\documentclass[12pt,a4paper]{article}
\usepackage{geometry}
\usepackage{graphicx}
\usepackage{color}
\newcommand{\sandor}[1]{#1}
\newcommand{\lstchr}[1]{\textbf{#1}}
\newcommand{\pred}{orange}
\begin{document}

\title{Full result for the QCD equation of state with 2+1 flavors}
\author{
Szabolcs~Bors\'{a}nyi$^a$,
Zolt\'{a}n~Fodor$^{a,b,c}$,
Christian~Hoelbling$^{a}$\\ 
S\'{a}ndor~D.~Katz$^{c,d}$,
Stefan Krieg$^{a,b}$,
K\'alm\'an~K.~Szab\'o$^{a,e}$\\
\ \\
$^a$Department of Physics, University of Wuppertal,\\ Gau\ss str. 20, D-42119, 
Germany\\
$^b$Forschungszentrum J\"ulich, J\"ulich, D-52425, Germany\\
$^c$Institute for Theoretical Physics, E\"otv\"os University,\\ P\'azm\'any
1, H-1117 Budapest, Hungary\\
$^d$MTA-ELTE Lend\"ulet Lattice Gauge Theory Research Group\\
$^e$Institute for Theoretical Physics, Universit\"at Regensburg\\
D-93040 Regensburg, Germany
}

\maketitle

\abstract{
We present a full result for the 2+1 flavor QCD equation of state. 
All the systematics are controlled, the quark masses are set to their physical values, and the continuum 
extrapolation is carried out. This extends our previous studies 
[JHEP 0601:089 (2006); 1011:077 (2010)] 
to even finer lattices and now includes ensembles with $N_t = 6,8,10,12$ 
up to $N_t = 16$. We use a Symanzik improved gauge and a stout-link improved staggered 
fermion action. Our findings confirm our earlier results.
In order to facilitate the direct use of our equation of state we make our
tabulated results available for download~\cite{resultsonline}.
}

\section{Introduction}
The early universe went through a rapid transition from a 
phase dominated by colored degrees of freedom to a phase dominated by 
color neutral degrees of freedom (hadrons). The same transition is now
routinely reproduced in heavy ion collisions at the Large Hadron Collider (LHC, 
CERN) and at the Relativistic Heavy Ion Collider (RHIC, Brookhaven 
National Lab.). The only systematic theoretical approach to determine 
many of the features of this transition is lattice QCD (for recent 
reviews see, e.g., \cite{Fodor:2009ax,Philipsen:2012nu,Petreczky:2013qj}).

Our most important qualitative knowledge about the transition is its 
nature. We know from lattice calculations that the transition 
is analytic. Therefore, without singular behavior, the system evolves smoothly 
from one phase to the other. (Since there is no real  
phase transition, the word ``phase'' merely indicates the dominant 
degrees of freedom.) This result~\cite{Aoki:2006we} of the 
Wuppertal-Budapest Collaboration was obtained through a finite size scaling 
analysis of the continuum extrapolated observables, computed using physical quark 
masses. Since there is a theoretically debated technical procedure\footnote{The so-called ``rooting'' trick, see~\cite{Fodor:2012gf} and references therein.} required in any $N_f=2+1$ flavor staggered calculation, as a future project it is 
desirable to reproduce the existing result using another -- preferably chiral -- lattice formalism.

One of the most important quantitative parameters of the transition is 
its absolute scale, the transition temperature $T_c$. In the absence of a 
real phase transition, there is, however, no uniquely defined
transition temperature.  
Different observables will lead to well defined $T_c$ values, which -- in principle -- 
can be determined to arbitrary precision. However, 
these $T_c$ values will likely not coincide (depending on the available 
precision). We obtained the first 
full\footnote{Here, by full result we mean results obtained 
using physical quark masses combined with a controlled continuum limit extrapolation, 
i.e. from ensembles with at least three lattice spacings in the 
scaling regime.} results for $T_c$ values for different 
observables in 2006~\cite{Aoki:2006br}.  
These results we later confirmed by simulations including successively finer and 
finer lattices \cite{Aoki:2009sc,Borsanyi:2010bp}. Furthermore, 
they were also confirmed by a recent independent calculation~\cite{Bazavov:2011nk}, thereby 
closing a long standing discrepancy. Depending on the exact 
definition of the observables, the remnant of the chiral transition 
(as we emphasized, there is no real phase transition, only an analytic 
``cross-over'') is at about $T_c=150$~MeV (for other observables see 
ref.~\cite{Borsanyi:2010bp}). Extending these results, the transition temperature 
was also determined for small non-vanishing baryonic chemical potentials ($\mu_B$) 
\cite{Endrodi:2011gv,Borsanyi:2012cr}. Here, we used the 
truncated version of the multiparameter-reweighting method of ref.~\cite{Fodor:2001au}. 
These (full) results provide the curvature of the phase diagram in the T-$\mu_B$ plane.

Describing the QCD transition and the phases below and above $T_c$ requires 
the determination of the equation of state (EoS). This means calculating 
the pressure ($p$), energy density ($\epsilon$), trace anomaly ($I=\epsilon-3p$), 
entropy ($s=(\epsilon+p)/T$) and the speed of sound ($c_s^2=dp/d\epsilon$) as functions of the temperature 
(and chemical potential). Several 
groups have determined the EoS with various methods, however, no full result 
(in the aforementioned sense) is available yet. A quite 
sensitive measure for the EoS is the peak height of the trace anomaly . Various 
approaches of the hotQCD Collaboration (p4, asqtad and hisq actions with 
$N_t$=6, 8, 10, and 12) resulted in a range of 5-8 for the peak ($I/T^4$, for a 
recent summary see ref.~\cite{Petreczky:2012fsa}). In 2005, the 
Wuppertal-Budapest group obtained a value slightly above 4, 
using two lattice spacings. This value was then confirmed
in the continuum limit~\cite{Borsanyi:2010cj}. Although 
\cite{Borsanyi:2010cj} provides full results for the EoS at three characteristic 
temperatures, no full result is available for the whole temperature 
range. Beyond the peak height, another disputed issue is the distance of the EoS from the 
Stefan-Boltzmann limit at, e.g., a temperature of $T=500$~MeV . 

The goal of this paper is to provide a full result for the $N_f=2+1$ EoS in a broad temperature 
range. Note that this also provides the missing piece of the $\mu_B>$0
equation of state  \cite{Borsanyi:2012cr}, where so far only 
the $\mu>0$ contribution could be quoted as a full result\footnote{The $\mu>0$ contribution to the pressure does not require renormalization. Therefore, a reliable continuum extrapolation for this part of the pressure was possible due to the lower computational costs than required for the $\mu=0$ part.}.
The present paper confirms our findings about the height of the peak, thus a resolution of 
the discrepancy remains to be a task for the future.

The outline of the paper can be summarized as follows. At first we describe 
our action and simulation setup. Then we present our analysis 
techniques, and finally the results are summarized and a conclusion is 
drawn. Since most of the readers are likely less interested in the lattice 
technicalities, but rather in the physics background and in the final 
results, we focus on these issues. Since the techniques employed are essentially 
the same as they were in our ref.~\cite{Borsanyi:2010cj}, we kindly refer the
reader to that paper (the only exception is the our histogram method 
\cite{Durr:2008zz}, which we will discuss in some detail). Since we do not 
determine the EoS for non-physical pion masses, the all-path method 
\cite{Borsanyi:2010cj,Endrodi:2010ai} is not used. For the 
practitioners we provide a table with our continuum results in ASCII format online~\cite{resultsonline}.

\section{Action, its physical motivation and simulation setup}
We use a tree-level Symanzik improved gauge action with 2-step 
stout-link improved staggered fermions. The precise definition of the 
action can be found in ref.~\cite{Aoki:2005vt}. Though the naive 
power counting tells us that the gauge part is more improved ($a^4$) 
than the fermionic part ($a^2$), this is no expensive overkill: 
the gauge part is relatively cheap and one wants to avoid a situation in 
which both sectors have the same order cutoff effects, but the 
computationally cheaper gauge part has accidentally a much larger 
prefactor. Indeed, our experiences show that using stout-link 
smearing the scaling features are very good~\cite{Aoki:2009sc}.

The main advantages of this action are threefold.

\lstchr{a.)} It is computationally fast (even faster than the completely 
unimproved action). This feature allows one to go to finer lattices 
(and thus closer to the continuum limit) at given computational
costs than with 
essentially any other action in the literature.

\lstchr{b.)} The action has a very well behaving continuum extrapolation. 
Our action approaches the continuum value of the 
Stefan-Boltzmann limit in the $T\rightarrow\infty$ limit slower than actions 
with p4 or Naik terms (the latter is an additional fermionic term in the asqtad 
and hisq actions). Nevertheless, our action is monotonous and reaches the 
asymptotic $a^2$ behavior quite ``early''. Extrapolations from moderate temporal 
extents, e.g., using $N_t \ge 8$, provide an accuracy on the percent level, 
which is the typical accuracy one aims to reach. Furthermore, stout link
smeared actions are ultralocal, 
both in fermion space as also in the gauge background, with a small exponential 
locality range in the gauge background only~\cite{Durr:2008zz}. According to experience, 
these features allow one to carry out a smooth continuum extrapolation. 
Furthermore, applying simple tree-level improvement factors for the bulk thermodynamic 
observables leads to results, which are already close to the continuum ones. 
Since the action is cheep to simulate, one can have several lattice spacings, which 
enables a controlled continuum extrapolation using many points. Other 
improved actions, with larger locality extent (p4 or Naik-type asqtad/HISQ) can 
have non-monotonic behavior (consider, e.g., the $N_t$ dependence of the free energy 
density for the Naik term)~\cite{Peikert:1997vf}. Furthermore, for $T=0$ simulations, 
required to compute the LCP and also to renormalize the $T>0$ data points, 
these actions have $O(a^2)$ cutoff effects. 
Therefore, their improvement which is motivated by 
$T\rightarrow\infty$ studies does not remove all $O(a^2)$ lattice artifacts at $T=0$.

\lstchr{c.)} Staggered fermions are cheap to simulate, however, the correct spectrum is 
recovered only in the continuum limit. In the 2+1 flavor framework 
there are 3/16 pseudo-Goldstone bosons instead of 3 (these are the pions 
in continuum QCD). To compensate for the deficit there is a tower of 
much heavier non-Goldstone pseudoscalars. As we approach the continuum 
limit, these heavier states merge with the 3/16 pseudo-Goldstone bosons 
and finally form the 3 pions that we need. This so-called 
``taste-violation'' at non-vanishing lattice spacing can be 
characterized by the splitting between the pseudo-Goldstone and the 
lowest level mass states in the pseudoscalar tower and also by the 
splittings within the tower. The smaller the splitting becomes
(i.e. the closer the non-Goldstone get to the pseudo-Goldstones bosons) 
the faster one reaches the continuum limit. This effect turned out to 
be more important for staggered QCD thermodynamics than improving the 
action in the $T\rightarrow\infty$ limit and motivated our choice 
of the two-stout smeared action for our large scale staggered QCD thermodynamics 
projects (see Figure~1 of ref.~\cite{Aoki:2005vt} or Figure~2 of ref.~\cite{Borsanyi:2010bp}). 
As of today, the new HISQ action possesses an even smaller taste violation (see, 
e.g., Figure 4 of ref.~\cite{Bazavov:2011nk}), though at higher computational costs.
Taste violation and/or heavier than physical quark masses increase the height of the trace
anomlay's peak. This fact is illustrated in Fig. 16 of Ref.~\cite{Borsanyi:2010cj}.

In the following four paragraphs we summarize the improvements 
over our previous EoS paper~\cite{Borsanyi:2010cj}.

\lstchr{a.)} In this paper the tree-level Symanzik improved gauge action with 
2-step stout-link improved staggered fermions is used on T$>$0 lattices 
with five lattice spacings corresponding to $N_t$=6, 8, 10, 12, and 16 
temporal extensions. The finest lattice $N_t$=16 is used to verify the 
earlier finding about the peak's height of the trace anomaly and to 
determine the additive renormalization (see below). This huge data set 
allows a fully controlled continuum extrapolation. In contrast, in 
ref.~\cite{Borsanyi:2010cj} we used $N_t$=6, 8, and 10 up to $T=$350~MeV above 
which only $N_t$=6 and 8 were used. At three characteristic temperatures 
also $N_t$=12 was included and the continuum extrapolation was carried 
out (with relatively large errors).

\lstchr{b.)} The T$\rightarrow$0 limit is particularly difficult to reach, since 
for a given $N_t$ lower and lower temperatures correspond to larger and 
larger lattice spacings, thus larger and larger taste violating 
effects. Nevertheless, this limit is important, since according to the 
standard choice, the renormalization is done at zero temperature: 
$p$($T$=0)=0. A mismatch at T=0 leads to a shift in the 
whole EoS. 
In ref.~\cite{Borsanyi:2010cj} we calculated the difference in the
pressure between the physical theory and its counterpart with 720~MeV heavy
pions at a selected temperature (100~MeV) on $N_t=6, 8$ and $10$ lattices.
At this low temperature
the latter theory has practically zero pressure, thus the difference
gives the pressure of the physical theory $p(T=100~\textrm{MeV})$ with the
desired normalization. 
Using this technique in ref.~\cite{Borsanyi:2010cj} we 
obtained about half (p/$T^4$=0.16) the value of the prediction of the 
hadron resonance gas model (p/$T^4$=0.27). This difference was included 
in the systematic error. Though this mismatch is really tiny compared to 
the Stefan-Boltzmann value of 5.209, it was obviously a suboptimal 
solution. Now we use all five lattice spacings (including  
$N_t$=16) to fix the additive term in the pressure, arriving at 
a complete agreement with the hadron resonance gas model at low 
temperatures.

\begin{figure}
\centerline{\includegraphics*[width=10cm]{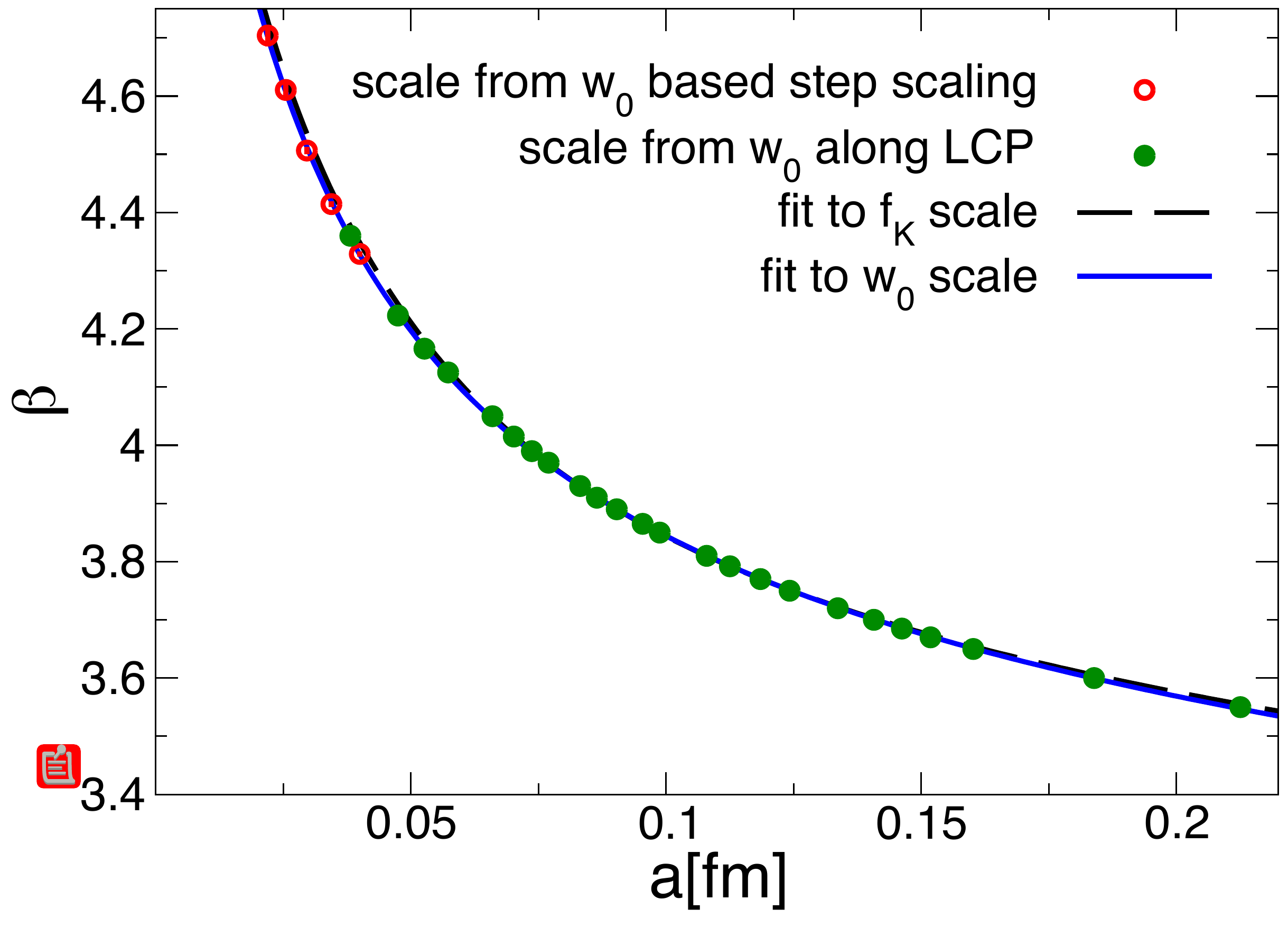}}
\caption{\label{scale}
The lattice spacing $a$ for the $\beta$ range used in this paper. Down to $a=0.047$~fm
two quantities were used to determine the scale, $f_K$ and $w_0$. 
For higher couplings a step-scaling approach was applied.
}
\end{figure}

\lstchr{c.)} We determined the scale and the line of constant physics (LCP) with
higher accuracy than in ref.~\cite{Borsanyi:2010cj}. For the scale 
in~\cite{Borsanyi:2010cj} a step-scaling technique was used
on lattice spacings smaller than 0.073~fm. 
Here we have a direct determination
of the lattice spacing based on the $w_0$ scale~\cite{Borsanyi:2012zs}
down to 0.047~fm.
Below that we again used a variant of
the step-scaling method~\cite{Luscher:1991wu} to determine the $\beta(a)$ function. To do this one needs a renormalized quantity which has a significant volume dependence even for very small physical volumes. We chose to use the derivative of the Yang-Mills flow of $E(t)=\langle F_{\mu\nu} F^{\mu\nu}\rangle$~\cite{Luscher:2010iy}
given by $t \cdot dt^2 E(t)/dt$. The flow time was coupled to the box size as $t=0.01L^2$, so our renormalized observable was $O=\left.t \cdot dt^2 E(t)/dt\right|_{t=0.01L^2}$.
In the first step
we used a physical volume corresponding to $L=24\cdot a_{\rm min}$ where $a_{\rm min}$ is the smallest lattice spacing we could reach with our direct approach.
Using this lattice and two coarser
ones, $16^4$ and $20^4$, both corresponding to the same physical volume, we extrapolated $O$ to a finer lattice spacing $a_1=24/32\cdot a_{\rm min}$.
Then, using simulations on $32^4$ lattices we searched for the $\beta_1$ coupling which provides this value of $O$. The pair ($\beta_1, a_1$) is the first new point of our scale function. In every further step this procedure was repeated: in the i-th step the physical box size was chosen as $L=24\cdot a_{i-1}$, the observable $O$ was extrapolated using $16^4$, $20^4$ and $24^4$ lattices to $a_i=24/32 a_{i-1}$ and the $32^4$ lattice was used to find $\beta_i$ which provides the extrapolated value of $O$ and thus corresponds to $a_i$.
The difference between the $f_K$ and $w_0$ scale settings is included into our 
systematic error estimate. 
We show the scale in Figure \ref{scale}. The LCP is defined by fixing the kaon decay 
constant to pion mass ratio $f_K/M_\pi$ and the light to strange quark 
mass ratio $m_s/m_{ud}$ to their physical values. (For the latter we 
used, similarly to our previous studies, a value of 28.15 as determined in 
ref.~\cite{Aoki:2009sc}. Most recent studies (e.g., refs. 
\cite{Durr:2010vn,Durr:2010aw}) lead to a 2\% lower value. Note that this 
is in the same ballpark as the accuracy of the LCP and, furthermore, this 
ratio has far less than this 2\% influence on the EoS. We studied this dependency 
of the EoS on the mass parameters in~\cite{Borsanyi:2010cj}.)

\lstchr{d.)} In ref.~\cite{Borsanyi:2010cj} we explicitly pointed out that ``for 
a rigorous continuum extrapolation one would need $N_t$=12 for the 
entire temperature region''. With the present paper we fulfill this 
condition. Furthermore, we extend our analysis procedure to control
the different sources of systematic errors, using our 
histogram method~\cite{Durr:2008zz}. We considered various fit methods 
(each of which is in principle a completely valid approach), 
calculated the goodness of fit Q and weights based on the Akaike 
information criterion AICc~\cite{Akaike}
 of that fit and looked at the unweighted or weighted (based on Q or AICc)
distribution of the results. The median gives 
our central value, whereas the central region containing 68\% of all the 
possible methods gives an estimate on the systematic uncertainties. This 
procedure provides very conservative errors. In the present case we had 
four basic types of continuum extrapolation methods (with or without tree level 
improvement for the pressure and with $a^2$ alone or $a^2$ and $a^4$ discretization 
effects) and two continuum extrapolation ranges 
(including or excluding the coarsest lattice $N_t$=6 in the analysis). 
We used seven ways to determine the subtraction term at T=0 (subtracting 
directly at the same gauge coupling $\beta$ or interpolating 
between the $\beta$ values with various orders of interpolation 
functions). We applied two scale setting procedures; one based on the kaon 
decay constant and one, as described above, using the $w_0$ scale. 
Finally, we had eight options to determine the 
final trace anomaly by choosing among various spline functions. This 
gives altogether 4$\cdot$2$\cdot$7$\cdot$2$\cdot$8=896 methods. Note that
using either an AICc or Q based distribution changed the result only by a tiny
fraction of the systematic uncertainty. Furthermore, the unweighted distribution
always delivered consistent results within systematical errors.

The systematic error procedure clearly 
demonstrates the robustness of our final result. Even in the case of 
applying or not applying tree level improvement, where the data points 
at finite lattice spacing change considerably, the agreement between the 
continuum extrapolated results, and hence the contribution to the 
systematic error, is on the few percent level.

\section{Results}
\begin{figure}
\centerline{\includegraphics*[width=15cm]{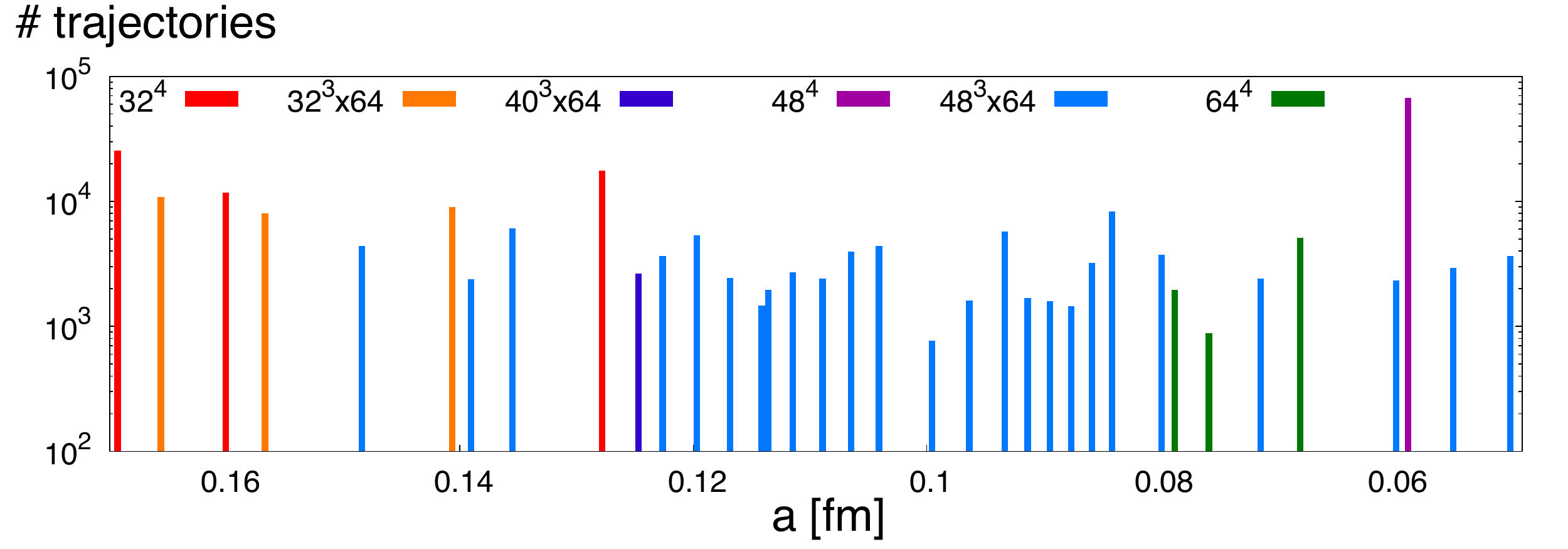}}
\caption{\label{coldstat}
The $T=0$ lattices and the collected statistics at various lattice spacings.
}
\end{figure}
Figure~\ref{coldstat} shows the lattice extents and the collected statistics
for our $T=0$ runs. Asymmetric ones were used for the scale setting, 
symmetric lattices for renormalization (and, additionally, for $w_0$ scale setting). 
We had the highest number of trajectories (67k) for the $48^4$ lattice which was used
to renormalize the $N_t=16$ data at $T=214$ MeV.
At finite temperature essentially
the same ensembles were used as in ref.~\cite{Borsanyi:2013hza},
with additional simulations on $32^3\times6$, $32^3\times8$ lattices
and $\sim$ 13$\times 10^3$ or 50$\times 10^3$ trajectories, respectively.
To reduce the potential finite-volume effects we also added six ensembles of
$48^3\times 12$ lattices in the range $T=220\ldots335$~MeV 
with $\sim 3\cdot 10^4$
trajectories, each.

At high temperatures we face two technical challenges. 

\lstchr{a.)} If the lattice
geometry is kept constant the physical volume will drop and relevant scales
might be absent from the lattice. To prevent this, we increased the volumes once more and generated ensembles with lattice sizes 
$32^3\times 6$, $48^3\times 8$, $64^3\times 10$, and $64^3\times 12$,
with 5, 40, 10 and 12 $\times 10^3$ trajectories, respectively.

\lstchr{b.)} The renormalization runs at high temperatures require extremely
fine lattices (below 0.05~fm lattice spacing), 
where $T=0$ simulations are no longer feasible
due to algorithmic limitations. Simulations are, however, still possible in the
deconfined phase. Starting from a temperature of $335$~MeV, we follow
the strategy described in refs.~\cite{Borsanyi:2012ve, Endrodi:2007tq}. 
Firstly, we subtract the value of the trace anomaly at the same coupling but 
doubled time extent (and thus a temperature of $T/2$ instead of $T$=0),
i.e.
$\left.(\varepsilon-3p)\right|_{T}
-\left.(\varepsilon-3p)\right|_{T/2}$. Adding to this result the value of the
trace anomaly at $T/2$ and the same $N_t$, we get the total trace anomaly.
For the half-temperature subtractions we generated ensembles on $48^3\times
16$, $64^3\times 20$ and $64^3\times 24$ lattices with matching parameters
and statistics to their finite temperature counterparts.

\begin{figure}
\begin{center}
\includegraphics*[width=8cm]{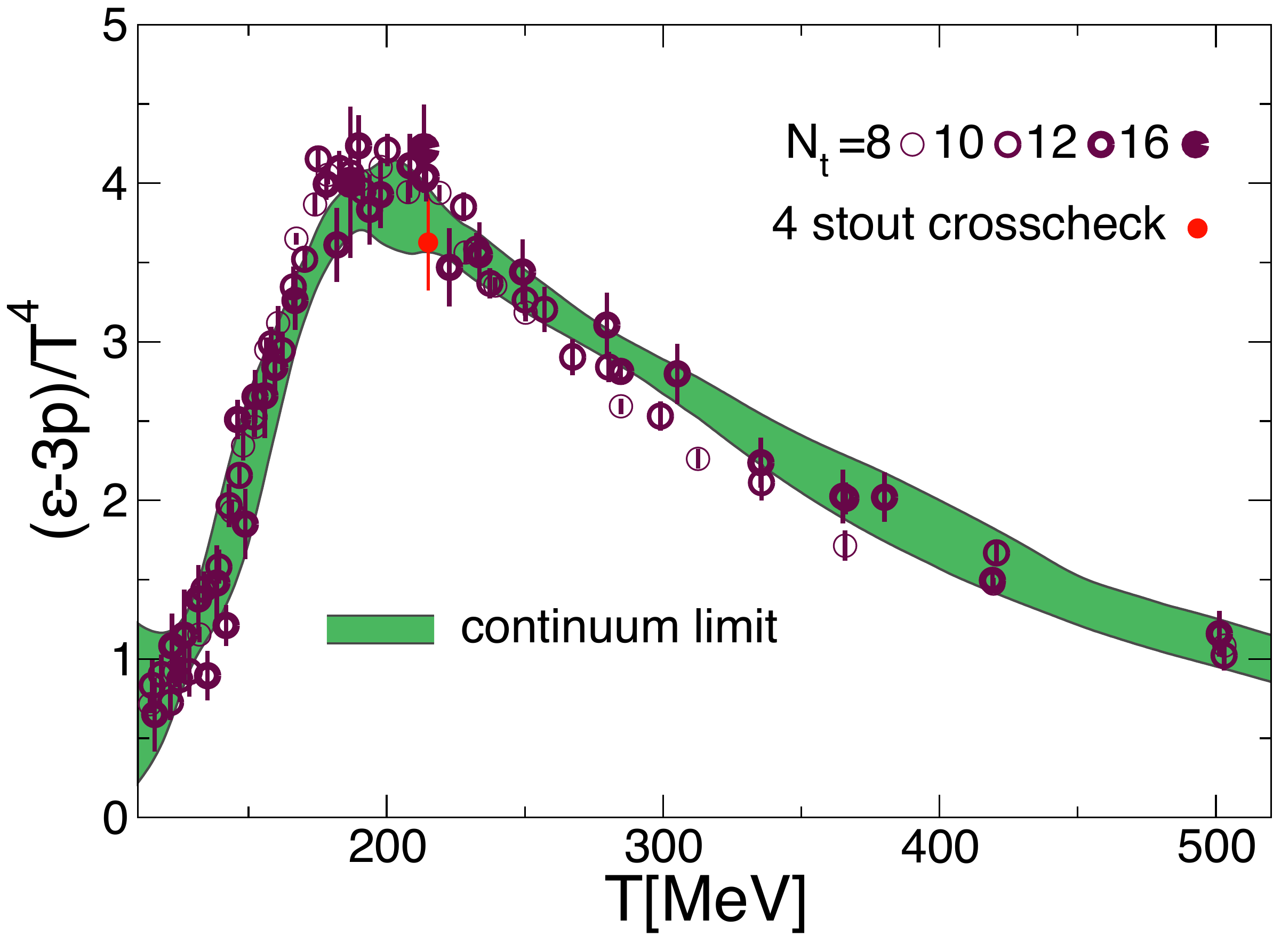}
\hspace{0.3cm}
\includegraphics*[width=8cm]{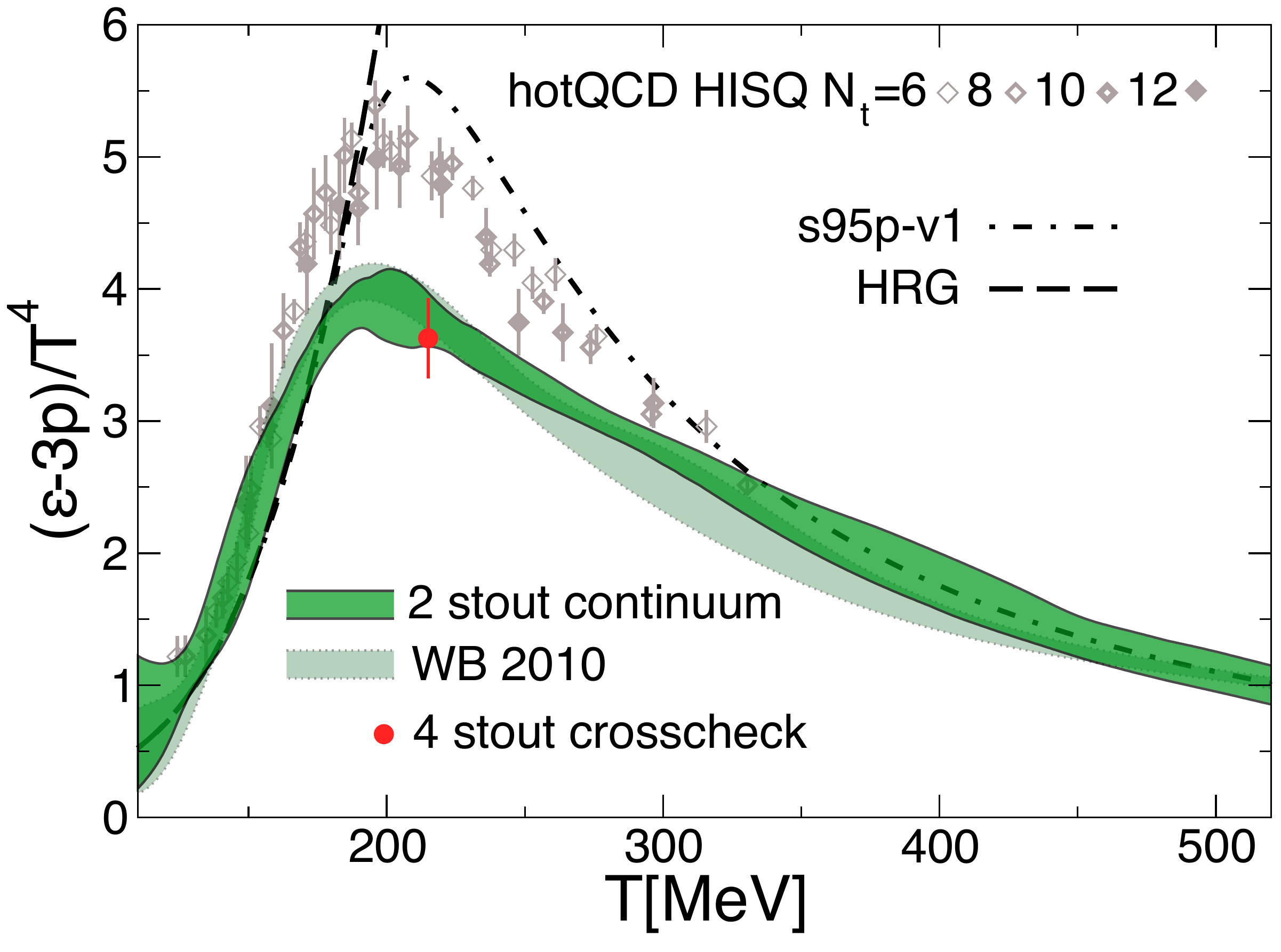}
\end{center}
\caption{\label{tracea_final}
\textit{Left:}
the trace anomaly as a function of the temperature for $N_t=8, 10, 12$ and $16$ lattices. The continuum extrapolated result including all systematic uncertainties is shown by the shaded band. Using a different action (see text), we performed a continuum extrapolation at the fixed temperature value of $214$~MeV, indicated here with a smaller filled red point. This independent result serves as a crosscheck on the peak's hight (also on r.h.s.). 
\textit{Right:}
comparison of the result with the parallel effort using the HISQ action by the hotQCD
collaboration (as it was presented at the Lattice 2012 conference
\cite{Petreczky:2012gi}, with $f_K$ scale setting) and the related parametrization 's95p-v1' of~\cite{Huovinen:2009yb}. A comparison to the Hadron Resonance Gas model's prediction and our result~\cite{Borsanyi:2010cj} from 2010 (``WB 2010'') is also shown.  }
\end{figure}

The continuum extrapolated trace anomaly is shown in Figure~\ref{tracea_final} (left).
In parallel to our investigations the hotQCD group is pursuing
a similar strategy to calculate the continuum extrapolated equation of state
using the HISQ action. As of the lattice
conference 2012 the results appear to be inconsistent~\cite{Petreczky:2012gi}. The situation might improve, however, when the HISQ analysis becomes complete with physical quark masses, a continuum extrapolation and a systematic error estimate.

\begin{figure}
\centerline{\includegraphics*[width=15cm]{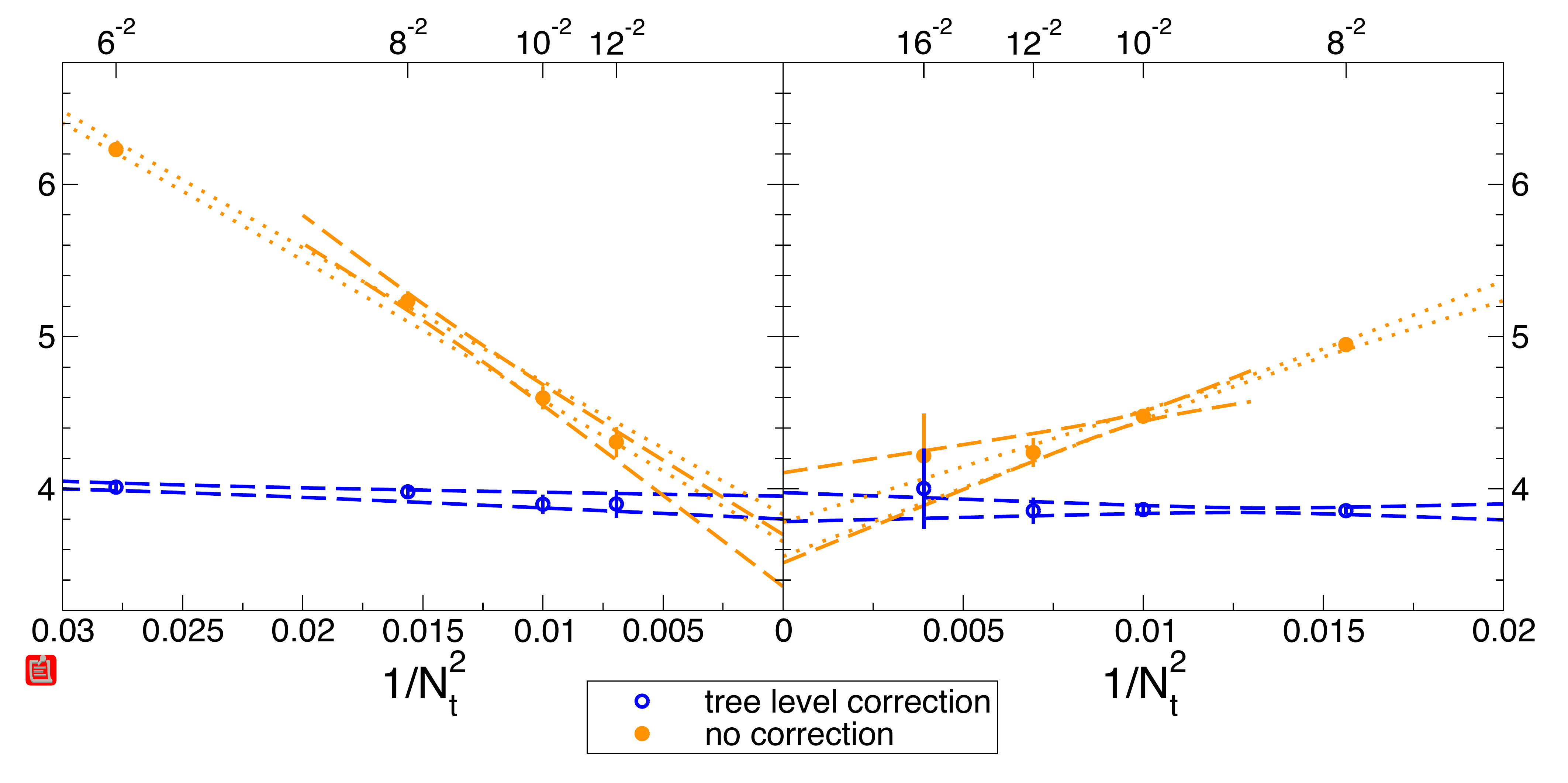}}
\caption{\label{cont_dual}
Continuum extrapolation of the trace anomaly at $T\approx 214$~MeV
with (blue points) and without (\pred points) tree level improvement. 
The left panel shows our results with $2+1+1$ flavors of 4-step 
stout improved staggered quarks extrapolating from 
$N_t=6,8,10$ and $12$,
whereas on the right panel $N_f=2+1$ 
2-step stout improved quarks are used on $N_t=8,10,12$ and $16$.
}
\end{figure}

The apparent discrepancy in Figure~\ref{tracea_final} is strongest
in the peak region between 200 and 230 MeV, thus
we have selected a temperature ($T\approx
214$~MeV) where we use an $N_t=16$ data point to demonstrate the continuum
scaling for our action. On the r.h.s. of Figure~\ref{cont_dual} we give the trace
anomaly both with and without tree level improvement. 
The extrapolated values are consistent with each other, 
however, tree level improvement results in smaller uncertainties. 

An ongoing project of the Wuppertal-Budapest collaboration is the determination
of the equation of state with $N_f=2+1+1$ flavors, i.e.~including
the contribution of the charm quark. This is done using a different action
with 4 steps of stout smearing with taste breaking artifacts roughly equal to
the HISQ action. 
Our 4-stout action has been tuned independently from our previous efforts 
by bracketing the physical point to $\pm 2\%$ in the quark masses, in boxes 
with $Lm_{\pi}>4$. The scale was set using the pion decay constant $f_\pi=130.41$~MeV.
 The tuning strategy was pursued to sub-percent accuracy down to a lattice spacing
  of $a\approx0.077$~fm, which corresponds to 
  $T\approx 214$~MeV temperature at $N_t=12$. 
Since the contribution of 
the charm quark is expected to be far below the present accuracy at $T=214$~MeV~\cite{Borsanyi:2012vn}, 
a continuum limit result for the trace anomaly at this temperature, 
computed using this new action, provides an independent 
cross-check of our 2-stout results. Therefore, in order to verify
our 2+1 flavor continuum extrapolation against a different action and scale
setting technique we made dedicated simulations using this 4-stout action 
at $T=214$~MeV. To also check for the volume dependence here we use the aspect ratio $LT=4$.
The result for the trace anomaly is shown in on the r.h.s of 
Figure~\ref{cont_dual}. Lattices up to $N_t=12$ are used. 
Both with and without tree level improvement, the continuum extrapolated trace
anomaly is in perfect agreement with that of the 2-stout action. This
cross-check with an independent regularization provides confidence on the
reliability of our continuum extrapolation. 

The pressure is obtained via integration from the trace anomaly.
As discussed before, a crucial step is to have the correct additive 
normalization to satisfy the $p(T=0)=0$ condition. We used our $N_t=16$
dataset to improve on the normalization presented in 
ref.~\cite{Borsanyi:2010cj} in the following way.

\begin{figure}
\centerline{\includegraphics*[height=6cm]{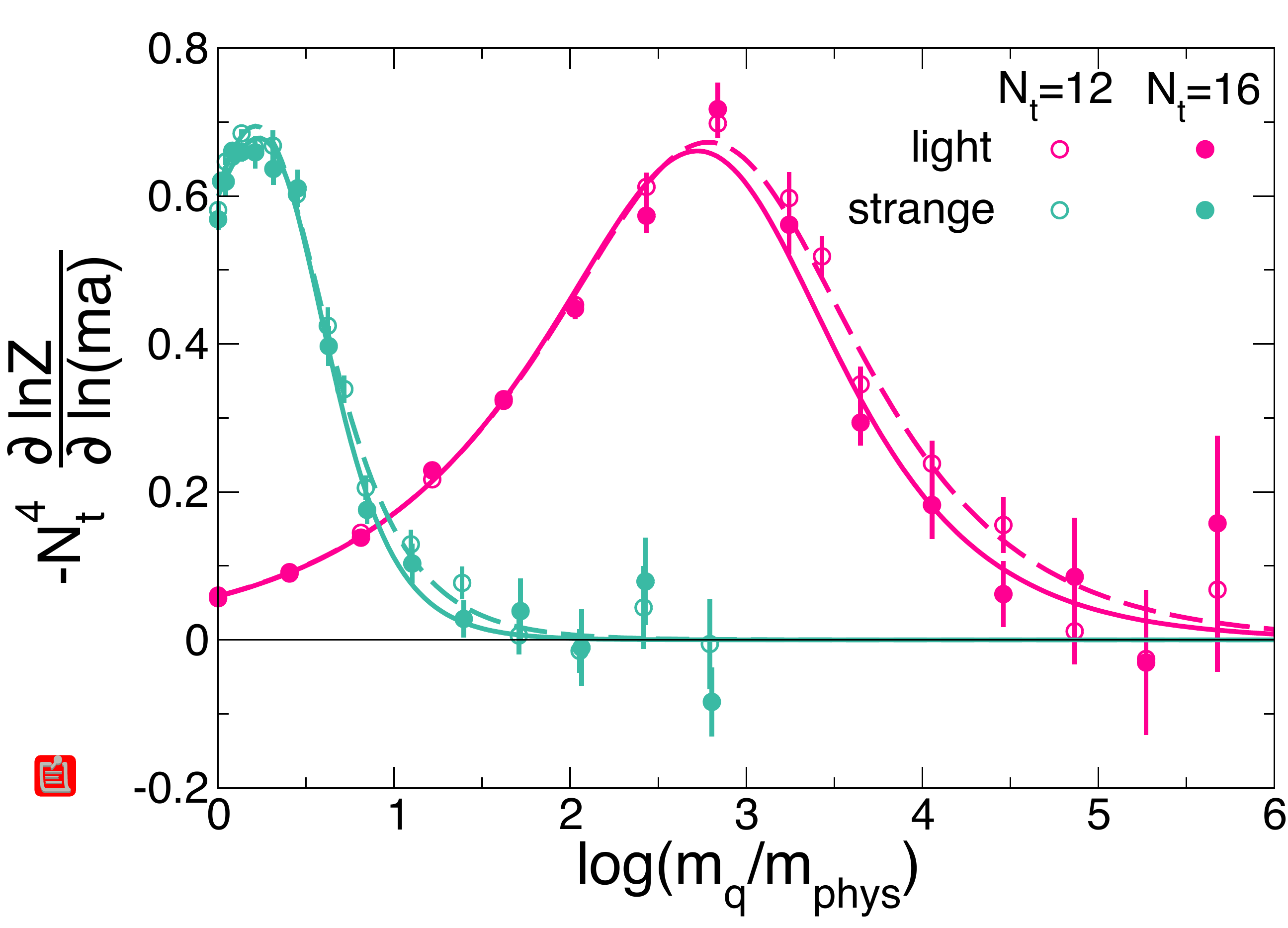}
\includegraphics*[height=6cm]{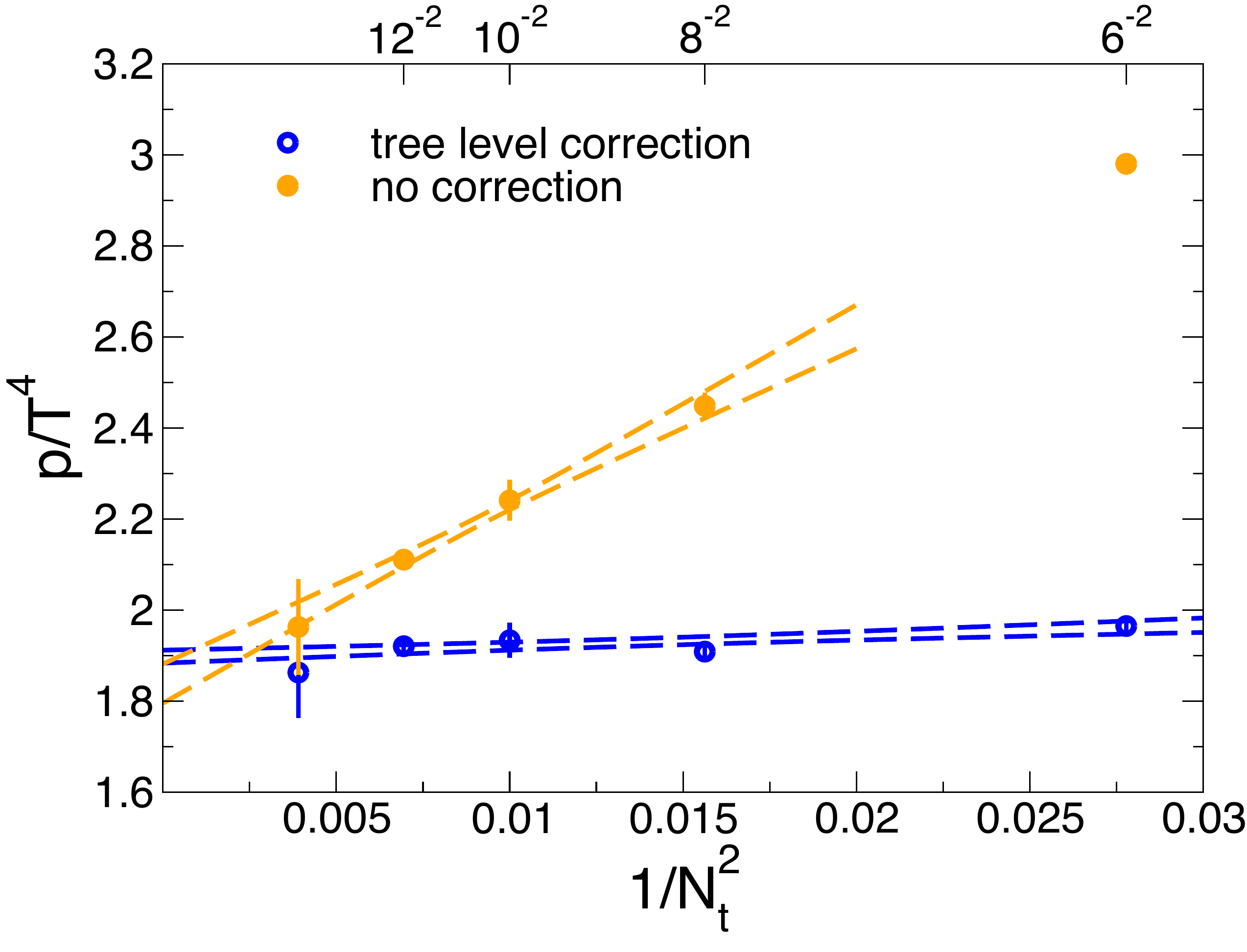}}
\caption{\label{normalization}
{\em Left:} contributions of the light (magenta) and strange quarks (turquoise) to the
the pressure at $T=214$~MeV at our two finest lattice spacings. The curves
represent a scan though various theories with different masses.
The sum of the area under the curves gives $p/T^4$.
{\em Right:}
continuum extrapolation of the pressure at $T\approx 214$~MeV with (blue)
and without (\pred) tree level improvement. Only statistical errors are shown.
}
\end{figure}

As it has also been discussed in ref.~\cite{Borsanyi:2010cj} the 
normalized pressure is directly related to the partition function
through $p/T^4=N_t^3/N_s^3\log Z$ in the thermodynamic limit.
Since $\log Z$ itself is not an observable one calculates derivatives
instead, such as the trace anomaly which can then be integrated to
give the pressure. There are many other possible choices for derivatives,
e.g., the bare mass parameter, which is equal to the chiral condensate.
\sandor{In order to get the correct additive renormalization for the pressure one
has to integrate from a starting point where the pressure
is equal to the $T=0$ pressure. We chose to integrate in the quark masses
along fixed gauge couplings, selected, for each $N_t$,
such that they correspond to $T_*=214$~MeV at the physical point. 
These same gauge couplings give a temperature that is deep in the confined phase for 
inifinitely heavy quark masses. Therefore integrating down from 
sufficiently heavy quark masses along these fixed couplings one gets the
correctly normalized pressure at $T_*=214$~MeV. The pressure at all other
temperatures is normalized to this point.}
In Figure~\ref{normalization} we show the derivatives with respect to the light
and strange quark masses as we vary the quark masses on a logarithmic
scale. Each magenta data point represent the chiral condensate coming from two
simulations, a finite temperature run and a dedicated renormalization run.
Here sea and valence quark masses are kept equal. For the same pair of runs
we show the strange contribution in turquoise.

We used this continuum result $p(T^*)$ in Figure~\ref{normalization} as
a starting point of the trace anomaly integration: 
\begin{equation}
\int^T_{T_*} \frac{\varepsilon(T')-3p(T')}{{T'}^5} dT'= 
\frac{p(T)}{T^4}-\frac{p(T_*)}{T_*^4}
\end{equation}

\begin{figure}
\begin{center}
\includegraphics*[width=8cm,height=6cm]{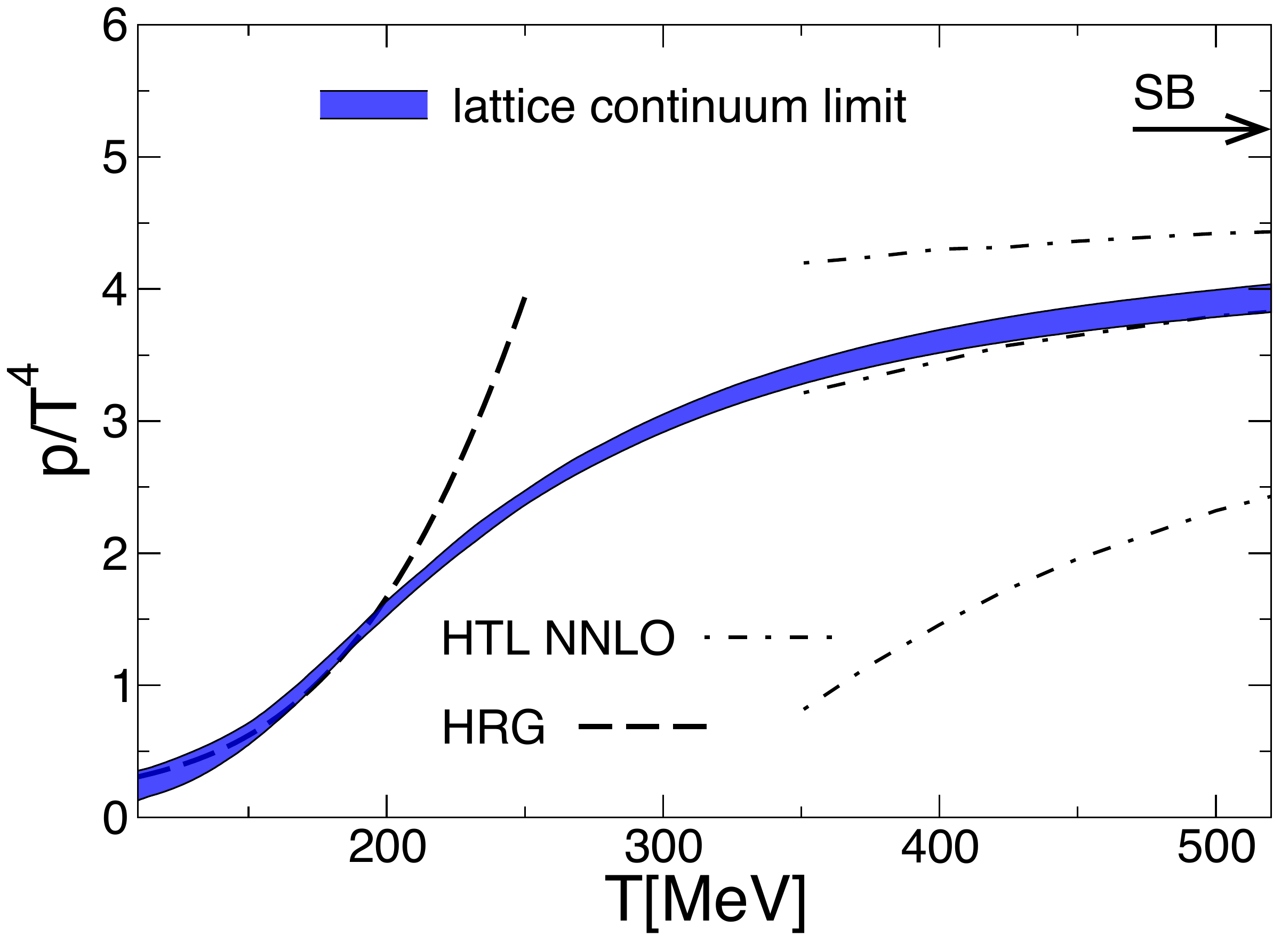}
\hspace{0.3cm}
\includegraphics*[width=8cm,height=6cm]{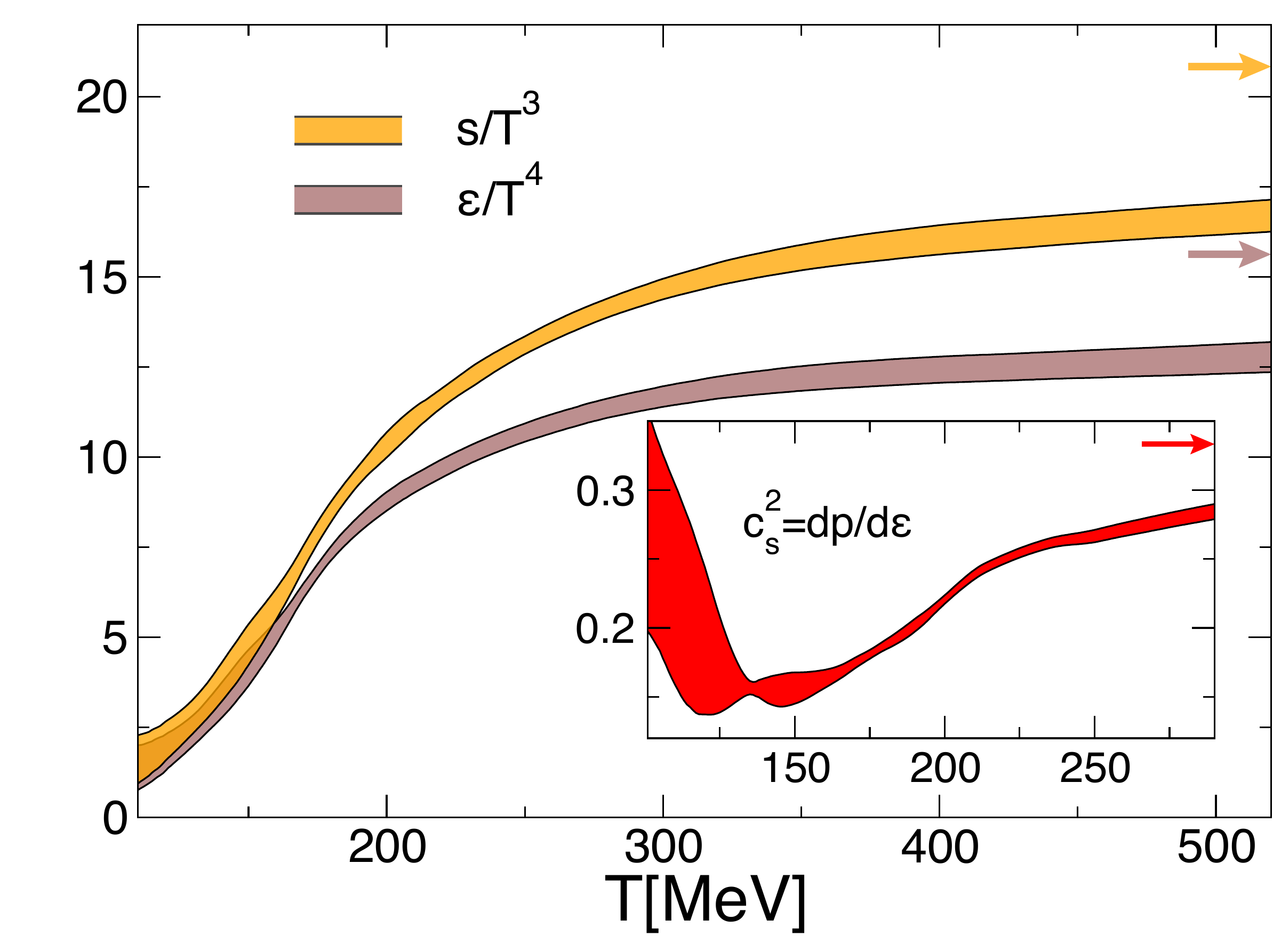}
\end{center}
\caption{\label{pressure}
{\em Left:}
continuum extrapolated result for the pressure with $N_f=2+1$ flavors.
The HRG prediction is indicated by the black line at low temperatures,
at high temperature we show a comparison to the NNLO Hard Thermal Loop
result of ref.~\cite{Andersen:2011sf} using three different renormalization
scales ($\mu=$$\pi T$, $2\pi T$ or $4\pi T$).
{\em Right:} entropy and energy density. The insert shows the 
speed of sound.
}
\end{figure}

The pressure is plotted in Figure~\ref{pressure} (left) together with the
predictions of the hadron resonance gas (HRG) model at low temperatures. There
is a perfect agreement with HRG in the hadronic phase. The energy and entropy
densities as well as the speed of sound are shown in the right panel of Figure~\ref{pressure}.

There is good agreement with our earlier result 
\cite{Borsanyi:2010cj}. The only obvious difference is that in the high 
temperature region (T$>$350~MeV), where we used only two lattice 
spacings previously (corresponding to $N_t$=6 and 8) the $N_t$=10 and 12 data 
changed the EoS by a few percent (the difference is 
on the 1.2 sigma level). This difference means 
that we should provide a new parametrization for 
the trace anomaly. The analytic function we suggest to use is of 
the same form:

\begin{equation}\label{fit_function} 
\frac{I(T)}{T^4} = \exp(-h_1/t-h_2/t^2)\cdot\left( h_0 + \frac{f_0 [ \cdot \tanh(f_1\cdot t+f_2)+1]}{1+g_1\cdot t + g_2 \cdot 
t^2} \right), 
\end{equation}
with the parameters of the fit are slightly changed. 
Table~\ref{fit_parameters} contain the parametrization of ref. 
\cite{Borsanyi:2010cj} and the parametrization of our present result. 
The maximal difference between our old parametrization~\cite{Borsanyi:2010cj} and the new 
one is only 2.8\% of the Stefan-Boltzmann value for the 
energy density. 
Note that though the two results differ only on the percent level,
the parameters in the new parametrization changed more (these changes
merely reflect some flat directions in the parameter space).

\begin{table}
\begin{center}
\begin{tabular}{|c||c|c|c|c|c|c|c|c|}
\hline
&$h_0$&$h_1$&$h_2$&$f_0$&$f_1$&$f_2$&$g_1$&$g_2$\\
\hline
\textbf{this work} & 0.1396 & -0.1800 & 0.0350 & 1.05 & 6.39 & -4.72 & -0.92 &0.57 \\
\textbf{2010~\cite{Borsanyi:2010cj}}& 0.1396 & -0.1800 & 0.0350 & 2.76 & 6.79 & -5.29 & -0.47 &1.04 \\
\hline
\end{tabular}
\end{center}

\caption{\label{fit_parameters}
Constants for our parametrization of the trace anomaly in Eq.~(\ref{fit_function}).
}
\end{table}

\section{Conclusions}
We have presented a full result for the $N_f=2+1$ QCD equation of state. Our
contiuum extrapolated results are completely consistent with our 
previous continuum estimate based on coarser 
lattices. The main advancement of the present work is the complete control
over all systematic uncertainties. We presented a parametrization
of our result which makes it easy to use in other calculations and provide 
our tabulated results for download.

\section*{Acknowledgments} Computations were performed on the Blue Gene 
supercomputer at FZ J\"ulich and on the QPACE machine and on GPU clusters 
\cite{Egri:2006zm} at University of Wuppertal. 
We acknowledge PRACE for awarding us resources on JUQUEEN at FZ J\"ulich.
CH wants to thank Utku~Can for interesting discussions.
This work was partially supported by the DFG Grant SFB/TRR 55 and ERC no. 208740.

\end{document}